\documentclass[aps,prl,showpacs,twocolumn]{revtex4}
\usepackage{graphics}
\usepackage{dcolumn}
\usepackage{bm}
\begin{document}
\title{ $\Lambda N$ space-exchange correlation effects in $_\Lambda^5$He 
hypernucleus}
\author{A. A. Usmani}
\email{ pht25aau@amu.ac.in}
\affiliation
{Department of Physics, Aligarh Muslim University, Aligarh 202 002, India}
\affiliation
{Inter University Centre for Astronomy and Astrophysics (IUCAA), Ganeshkhind, 
Pune-411 007, India}

\date{\today}
\begin{abstract}
A complete realistic study of $_\Lambda^5$He hypernucleus is presented 
using a realistic Hamiltonian and wave function. This study takes into account
all relevant dynamical correlations along with $\Lambda N$ space-exchange 
correlation ($SEC$). We also compute $\Lambda NN$ force and the correlation 
induced by this force.  The $SEC$ affects the central repulsive $\Lambda N$
correlation significantly at $r\le 2.0$ fm, specially at its peak and in its
vicinity. $SEC$  significantly affects energy breakdown of the hypernucleus, 
$\Lambda$-seperation energy, nuclear core ($NC$) ploarization and 
density profiles.  A large $NC$ polarization is found with and without
$SEC$, respectively. 
The $SEC$ effect is relatively large in two-pion exchange component of
$\Lambda NN$ force. Therefore, any attempt to pin down the strength of this
force with no $SEC$ would be flawed.
\end{abstract}


\maketitle
Very recently, realistic studies \cite{AUsmani03,RSinha03} 
have been performed on s-shell single hypernuclei 
using realistic two-nucleon ($NN$) Argonne $v_{18}$ \cite{AV18} potential and  
three-nucleon ($NNN$) Urbana model-IX \cite{NNN-IX,NNN} potential in 
the non-strange 
sector in conjunction with the two-baryon ($\Lambda N$) Urbana type charge 
symmetric potential \cite{BUC84,Lagaris81} and three-baryon ($\Lambda NN$) 
potential \cite{Bhaduri67,BU88,Gal75} in the strange sector. 
In an alternative approach, Nemura {\it et.al.} \cite{HNemura02}
have also performed an {\it ab-initio} calculation on all the s-shell
hypernuclei by explicitily including $\Sigma$ degree of freedom
at the two-body level. 
Besides, there have been studies of $_\Lambda^5$He \cite{AUsmani95} and of 
$_{\:\:\Lambda}^{17}$O \cite{UPU95} using truncated NN ($v_6$) potential. 
A couple of these studies were aimed to pin down the 
strengths of $\Lambda NN$ force \cite{RSinha03,AUsmani95}.
In the above as well as in other realistic, microscopic 
studies of hypernuclei \cite{BU88,DHT72,QUsmani80}
to date, the $\Lambda N$ space-exchange correlation ($SEC$)
has always been put aside while writing the wave function that describes the
hypernucleus. This is despite the fact that the expectation value of 
the corresponding $\Lambda N$ space-exchange potential 
which arises due to an equivalent $\Lambda N$ interaction in the 
relative p-state is not small.  
This has been demonstrated in the 
Urbana charge symmetric potentials, in many such recent 
calculations. 

Having  $\Lambda N$ potential 
\begin{equation}
\label{vLambdaN}
v_{\Lambda N}(r)=v_0(r) (1-\varepsilon+\varepsilon P_{x}^{\ell})
+\frac{1}{4}v_{\sigma}T_{\pi}^{2}
\mbox{\boldmath$\sigma$}_{\Lambda}\cdot\mbox{\boldmath$\sigma$}_{N},
\end{equation}
\begin{equation}
v_0(r)=\frac{W_c}{1+{\rm exp}(r-R)/a}-\overline{v}T_{\pi}^{2}.
\end{equation}
one solves the Schr\"{o}dinger equation 
\begin{equation}
\label{Schrodinger}
\left[\frac{-\hbar^{2}}{2\mu_{\Lambda N}}\nabla^{2}
+\tilde{v}^{\ell}_{s(t)}(r)
+\theta_{\Lambda N}
+\frac{\hbar^{2}\ell (\ell+1)}{2\mu_{\Lambda N} r^2}
\right]f_{s(t)}^{\ell}(r)=0,
\end{equation}
to obtain the radial solutions $f^\ell_s(r)$ and $f^\ell_t(r)$ with the help 
of the quenched $\Lambda N$ potentials in singlet and triplet states
 \begin{equation}
\label{alpha1}
\tilde{v}^{\ell}_{s}(r)=[v_{c}(r)-\alpha_{2\pi}\bar{v}T_\pi^2]
(1-\varepsilon+\varepsilon P_x^{\ell})
+\frac{3}{4} \alpha_{\sigma}v_{\sigma}T^2_{\pi}, 
\end{equation}
\begin{equation}
\label{alpha2}
\tilde{v}^{\ell}_{t}(r)=[v_{c}(r)-\alpha_{2\pi}\bar{v}T_\pi^2]
(1-\varepsilon+\varepsilon P_x^{\ell})
-\frac{1}{4} \alpha_{\sigma}v_{\sigma}T^2_{\pi}.
\end{equation}
where $P^{\ell}_x$ is a Majorana space-exchange operator and 
$\varepsilon$ is the corresponding exchange parameter.  
The $\overline{v}=(v_{s}+3v_{t})/4$ and $v_{\sigma}=v_{s}-v_{t}$ are, 
respectively the spin-average and spin-dependent strengths, 
with $v_{s(t)}$ the singlet(triplet) state depths. 
$T_\pi$ is the one-pion exchange (OPE) tensor potential and
$\theta_{\Lambda N}$ is an auxiliary potential that ensures the asymptotic 
behaviour of long range correlation functions.

The central repulsive $\Lambda N$ correlation has a radial dependence
\begin{equation}
\label{f_c}
f^c_{\Lambda N}(r)=f_{\Lambda N}^{0}(r). 
\end{equation}
With the $SEC$ having radial dependence
\begin{equation}
\label{u^x}
u^x(r) =\frac{f^{0}_{\Lambda N}(r)-f^{1}_{\Lambda N}(r)}{2},
\end{equation}
central correlation is modified as described by 
\begin{equation}
\label{f_cx}
f^c_{\Lambda N}(r)
=\frac{f^{0}_{\Lambda N} (r)+f^{1}_{\Lambda N}(r)}{2}
=f^{0}_{\Lambda N}(r) -u^x(r).
\end{equation}
Here $f^{\ell}_{\Lambda N}(r)$ is the spin averaged 
correlation function,
\begin{equation}
f^{\ell}_{\Lambda N} (r)=\frac{f^{\ell}_s (r)+3f^{\ell}_t (r)}{4}.
\end{equation}
The weak spin-spin correlation function
\begin{equation}
u^{\sigma}_{\Lambda N} (r)=\frac{f^{0}_s (r)-f^{0}_t (r)}{4}
\end{equation}
is found insignificant in this study.

In Ref. \cite{AUsmani03} the variational wave function \cite{Arriaga95} for 
nuclei  is  generalized: 
\begin{eqnarray}
\label{Psi}
\mid\Psi\rangle &=& 
\left[1+\sum_{j<k}^{A-1}U_{\Lambda jk}
+\sum_{i<j<k}^{A-1}(U_{ijk}+U_{ijk}^{TNI})\right.\nonumber \\
&&\left.+\sum_{i<j}^{A-1} U_{ij}^{LS}\right] \nonumber  \mid \Psi_p\rangle, 
\end{eqnarray}
where the pair wave function, $\Psi_p$, is 
\begin{equation}
\label{Psi_p}
\mid\Psi_p\rangle=\left[\prod_{j=1}^{A-1}(1+U_{\Lambda j})\right]
\left[S\prod_{i<j}^{A-1}(1+U_{ij})\right]
\mid \Psi_{J}\rangle.
\end{equation}
Here $U_{ij}$, $U_{ij}^{LS}$, $U_{\Lambda j}$, $U_{ijk}$, $U_{ijk}^{TNI}$ 
and $U_{\Lambda jk}$ are the non 
commuting two- and three- baryon correlation operators.
$S$ is the  symmetrization operator and
$\mid \Psi_J\rangle$ is the antisymmetric Jastrow wave function
\begin{eqnarray}
\label{Psi_J}
\mid \Psi_{J}\rangle &=& 
\left[\prod _{j<k}^{A-1} f^{c}_{\Lambda jk}\right] 
\left[\prod _{j=1}^{A-1} f^{c}_{\Lambda}(r_{\Lambda j})\right] 
\left[\prod _{i<j<k}^{A-1}f^{c}_{ijk}\right] 
\nonumber \\
&&\times\left[\prod _{i<j}^{A-1}f^{c}_{ij}(r_{ij})\right] 
{\cal A}\mid \phi^{A-1}\rangle.
\end{eqnarray}

Using variational Monte Carlo (VMC) method we calculate the 
$\Lambda$-seperation energy
\begin{equation}
\label{BLambda}
B_\Lambda=
\frac{\langle \Psi_{A-1}|H_{NC}|\Psi_{A-1}\rangle}{\langle\Psi_{A-1}|
\Psi_{A-1}\rangle}
-\frac{\langle \Psi_A|H|\Psi_A\rangle}{\langle\Psi_A|\Psi_A\rangle},
\end{equation}
where  $\Psi_A$ and $\Psi_{A-1}$ are the wavefunctions of the hypernucleus 
and its isolated bound $NC$. 
Similarly, $H$ and $H_{NC}$ are used to refer to the
non-relativistic Hamiltonians of the hypernucleus and its $NC$. 

The Hamiltonians are written using potentials as mentioned.
For the $\Lambda NN$ force, one may write two Wigner types of forces, namely,
the dispersive $\Lambda NN$ force ($V_{\Lambda NN}^D$) 
suggested by the suppression 
mechanism due to $\Lambda N-\Sigma N$ coupling \cite{BR71,Rozynek79,BR70,UB99},
and the two-pion exchange ($TPE$) $\Lambda NN$ force  ($V_{\Lambda NN}^{2\pi}$)
\begin{equation}
V_{\Lambda NN}=V_{\Lambda NN}^{D}+V_{\Lambda NN}^{2\pi}.
\end{equation}
The phenomenological dispersive force with  
explicit spin dependence is written as \cite{BU88} 
\begin{equation}
\label{VLNND}
V_{\Lambda ij}^{D}=
W^{D}T_{\pi}^{2}(r_{\Lambda i})T_{\pi}^{2}(r_{\Lambda j})
  [1+\frac{1}{6}\mbox{\boldmath$\sigma$}_{\Lambda}\cdot(
\mbox{\boldmath$\sigma$} _{i}+ \mbox{\boldmath$\sigma$}_{j})].
\end{equation}
$V_{\Lambda NN}^{2\pi}$ is written as a sum of two terms 
due to p- and s-wave $\pi-N$ scatterings given below
\begin{equation}
\label{VPW}
V^{PW}_{\Lambda ij}=-\left(\frac{C^{PW}}{6}\right)
(\mbox{\boldmath$\tau$}_{i}\cdot
\mbox{\boldmath$\tau$}_{j})
\{X_{i\Lambda},X_{\Lambda j}\}
\end{equation}
and
\begin{equation}
\label{VSW}
V^{SW}_{\Lambda ij}=C^{SW}
Z(\mu  r_{i\Lambda})Z(\mu r_{j\Lambda })
\mbox{\boldmath$\sigma$}_i\cdot {\hat{\bf r}}_{i\Lambda }
\mbox{\boldmath$\sigma$}_j\cdot {\hat{\bf r}}_{j\Lambda}
\mbox{\boldmath$\tau$}_i\cdot\mbox{\boldmath$\tau$}_j,
\end{equation}
with
\begin{eqnarray}
&\label{XiL}
X_{\Lambda i}=( \mbox{\boldmath$\sigma$}_{\Lambda}\cdot  
\mbox{\boldmath$\sigma$}_{i} )Y_{\pi}(r_{\Lambda i})+S_{\Lambda i}
T_{\pi}(r_{\Lambda i}),\nonumber \\
&Z(x)=\frac{x}{3}\left[Y_\pi(x)-T_\pi(x)\right].
\end{eqnarray}
$W^D$, $C^{PW}$ and $C^{SW}$ are the strengths of the potential. 
$S_{\Lambda i}$ is a tensor operator, $Y_\pi$ is OPE Yukawa function 
and subscripts $i,j$ and $\Lambda$ refer to two nucleons 
and a $\Lambda$ in the triplet ($\Lambda ij$). $V^{SW}_{\Lambda NN}$
is discussed in detail in Ref. \cite{AUsmani04}.  
In order to obtain optimal realistic wave function 
we include all correlations
induced by these potential pieces in $U_{\Lambda ij}$ as \cite{AUsmani03}
\begin{equation}
\label{eq7}
U^{m}_{\Lambda ij}= \epsilon_{m=D,SW,PW} V^{m}
(\tilde{r}_{\Lambda i},\tilde{r}_{ij}, \tilde{r}_{j\Lambda}),
\end{equation}
using $\epsilon_m$ as a 
variational parameter and $\tilde{r}$ as a scaled $\Lambda N$ pair
distance used for triplet correlation function.

\begin{figure}
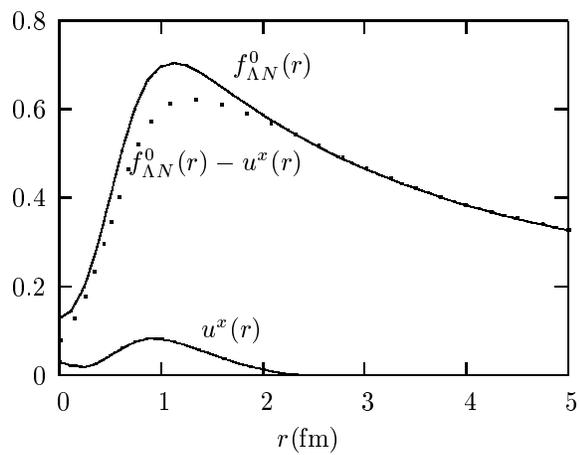

\caption{\label{figone}
$\Lambda N$ central correlation and space-exchange correlation 
$(SEC)$ functions.}
\end{figure}

\begin{table}
\caption{\label{tabone} Energy breakdown for $^{5}_{\Lambda}$He hypernucleus. 
All quantities are in units of MeV.}
\begin{ruledtabular}
\begin{tabular}{lccc}
\hline
 &  with $u^x$  & with no  $u^x$ & Difference \\
 &   A  & B &  A-B \\
\hline
 $T_{\Lambda}$        & 8.75(3)   & 9.05(3)  &-0.30(6)   \\
 $v_0(r)(1-\epsilon)$ & -12.56(4) & -13.06(4) & 0.50(8) \\
 $v_0(r)\epsilon P_x^0$ & -5.24(2)  & -4.79(2)  & -0.45(4) \\
 $\frac{1}{4}v_\sigma T_\pi^2 \mbox{\boldmath$\sigma$}_\Lambda\cdot\mbox{\boldmath$\sigma$}_N $ & 0.0   & 0.0  & 0.0\\
 $V_{\Lambda N}$  & -17.80(6)  & -17.85(6)   & 0.05(12) \\
 $V_{\Lambda NN} ^D$   & 2.38(1) & 2.50(1)  & -0.12(2) \\
 $V_{\Lambda NN}^{2\pi}$ & -2.46(2)  & -2.88(2) & 0.42(4) \\
 $V_{\Lambda NN}=V_{\Lambda NN}^D
   +V_{\Lambda NN}^{2\pi}$ & -0.13(1)   & -0.38(1)   & 0.25(2)  \\ 
 $V_\Lambda^{Total}=v_{\Lambda N}+V_{\Lambda NN}$ 
                       & -17.93(6)   &  -18.23(6)  & 0.30(2)    \\
 $E_\Lambda=T_{\Lambda}+ V_\Lambda^{Total}$
 &  -9.18(4)  &  -9.18(4)  & 0.0(8) \\
 $T_{NC}$     &  117.38(18)     &  116.58(18)  & 0.80(36) \\
 $v_{NN}$     & -133.90(16)    & -132.50(16)  & -1.40(32)  \\
 $V_{NNN}$    &  -5.67(2)      & -5.77(2)   &  0.10(4)\\
 $V_{NC}=v_{NN}+V_{NNN}$  & -139.57(16)  & -138.27(16)  & -1.30(32) \\
 $E_{NC}=T_{NC}+V_{NC}$  &  -22.19(4)      &  -21.68(4)  &-0.51(8)   \\
$E(_\Lambda^5{\rm He})$ &-31.40(2) & -30.85(2)  &-0.55(4) \\
 $B_{\Lambda}$( Expt. 3.12(2))   & 3.66(3)  & 3.11(3) & -0.55(4)  \\
\hline
\end{tabular}
\end{ruledtabular}
\end{table}

\begin{table*} 
\caption{\label{tabtwo}
$NC$ polarization. All quantities are in units of MeV. }
\begin{ruledtabular}
\begin{tabular}{lccccc}
\hline
&   & \multicolumn{2}{c}{ With $u^x$} &\multicolumn{2}{c}{ With no  $u^x$}   \\
&   $^4$He & $NC$ & Polarization & $NC$ & Polarization \\
 & A & B & B-A & C & C-A  \\
\hline
$T_{NC}^{internal}$ &107.94(14) &115.63(18) &7.69(7)&114.73(18)& 6.79(7)  \\
  $v_{NN}$  & -130.45(16) & -133.90(32) & -3.45 (6) & -132.50(16) & -2.05(6) \\
 $V_{NNN}$  & -5.21(2)  & -5.67(2)  & -0.46(2) & -5.77(2) & -0.56(2)  \\
$V_{NC}=v_{NN}+V_{NNN}$&-135.66(14)&-139.57(18)&-3.91(4)&-138.27(14)&-2.61(4) \\
$E_{NC}^{internal}=T_{NC}^{internal}+V_{NC}$&-27.73(1)&-23.95(2)&3.78(4)&-23.55(2)&4.18(4)\\
\hline
\end{tabular}
\end{ruledtabular}
\end{table*}

The space-exchange operation interchanges the positions of 
$\Lambda$ and nucleon in a  $\Lambda N$ pair and thereby, 
it affects the centre of mass (c.m.) 
of the hypernucleus.
This is redetermined after every such operation using
\begin{equation}
{\bf R}_{c.m.}= \frac{m\sum_{i=1}^{A-1} {\bf r}_i+m_\Lambda {\bf r}_\Lambda}
{m_N(A-1)+m_\Lambda}
\end{equation}
as all the positions of baryons are measured from the c.m. of the whole system, 
$\tilde{\bf r}={\bf r}-{\bf R_{c.m.}}$. This is essential to make the 
wave function translationally invariant.
Consequently, one has to exchange positions
of a  $\Lambda$ with all the  nucleons one by one, keep track of the 
shift in c.m. and distances due to this operation, and then repeatedly
calculate the wave function involving all other two-body and three-body 
correlations.
This is computationally very difficult to implement.

It turns out \cite{UPU95,AUsmani03} that the  strong repulsive 
central correlation has a major contribution
to binding energies, nuclear core ($NC$) polarization 
and density profiles of $N$ and $\Lambda$.
$NC$ polarization is just the rearrangement energy which is the 
difference of internal energy (defined later) of the $NC$ in the hypernucleus 
and the energy of an  identical isolated bound nucleus.
A significant modification of this correlation
may strongly affect the above results.
In Fig. \ref{figone}, we plot the repulsive central correlation
in case of no  $SEC$ given by Eq. (\ref{f_c}) and 
and also with $SEC$ given by Eq. (\ref{f_cx}) along with $u^x(r)$. 
These functions have been obtained using our best optimal variational
wave functions for both the cases. We note a substantial modification 
of central repulsive correlation at the peak of the correlation and in its
vicinity due to presence of  $u^x(r)$. The $u^x(r)$ is found substantially
effective at $r\le 2.0$ fm and has no effect at $r\ge 2.4$ fm 
where it vanishes,
thereby $f^c_{\Lambda N}(r)$ remains the same for both the cases. 
This does warrant a crucial role of $SEC$ in hypernuclear studies. 
Findings of this work suggest that a study ignoring $SEC$
would be misleading, specially while determining the strength of $\Lambda NN$ 
potential discussed below.  Any attempt to resolve the notorious $A=5$ 
anomaly \cite{DHT72,Gal75,BPovh80,Hungerford84} would be hopelessly defecient
if $SEC$ is ignored.

We present our energy results for the cases (i) with no $SEC$ (ii) and with 
$SEC$ in the wave function. For the first case, we optimize our variational
wave function and choose a value for 
the strength $C^{PW}$=0.75 MeV 
and then adjust the dispersive strength $W^D$ in order to reproduce the
experimental $\Lambda$-seperation energy $B_{\Lambda}=3.12(2)$ MeV. Doing so,
$W^D$ turns out to be 0.125 MeV. We then switch on the $SEC$ and again 
optimize the variational wavefunction and then calculate the energy of the 
hypernucleus. In this case, we note that the hypernucleus is over bound 
by 0.55(4) MeV.  Therefore, $B_\Lambda$ turns out to be 3.65(3) MeV.
The detailed energy breakdown is presented in Table \ref{tabone}. 
We note a significant contribution of $SEC$ in every piece of energy.
As reported \cite{UPU95,AUsmani95,AUsmani03,AUsmani04} 
$V^{2 \pi}_{\Lambda NN}$ has a generalised tensor-tau type structure. It is 
found sensitive to operatorial correlations as well as to $SEC$.
The contribution of $SEC$ to central $\Lambda N$ potential 
($v_0(r)(1-\varepsilon)$) is 0.50(8) MeV. It
is alomst cancelled by its contribution to space-exchange part of the potential
$(v_0(r)\varepsilon P_x)$ which is found to be -0.45(4) MeV. 
Its contribution to $\Lambda$ kinetic energy $T_\Lambda= -0.3$ MeV 
is balanced by its relatively large effect
on $V^{2\pi}_{\Lambda NN}$ which is found to be 0.25(2) MeV in which
$V_{\Lambda NN}^{2\pi}$ is 0.42(4) MeV and $V_{\Lambda NN}^D$ is -0.12(2) MeV.  
Therefore, $\Lambda$-energy turns out to be the
same i.e. -9.18(4) MeV with and with no $SEC$. This may be 
accidental in case of $_\Lambda^5$He that may not hold in general.
For both the cases we note significant difference of 0.55(4) MeV in 
the binding energy of hypernucleus as well as its $B_\Lambda$ value.  
This is due to the effect of $SEC$ in  $NC$ part of the energy,
which adds to $NC$ polarization or rearrangement energy.

\begin{figure}
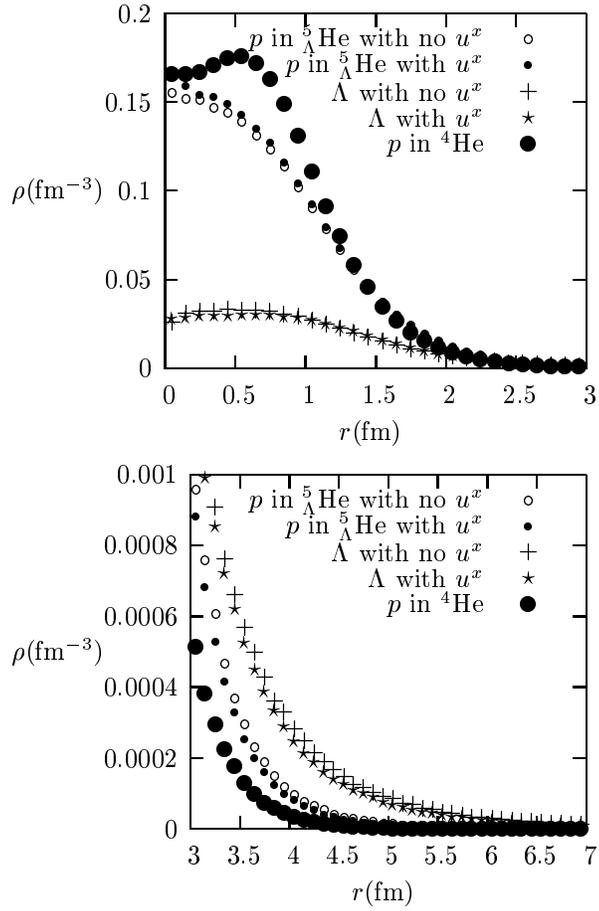

\caption{\label{figtwo}
Density profiles of $p$ and  $\Lambda$.  }
\end{figure}

In order to calculate the polarization energy of $NC$, we calculate the 
internal energy $(E_{NC}^{internal})$ of the $(A-1)$ subsystem. 
Taking into account the c.m. motion of the subsystem, we write down
\begin{equation}
T_{NC}^{internal}= \sum_{i=1}^{A-1} \frac {p_i^2}{2m} 
- \frac {\left(\sum_{i=1}^{A-1} p_i\right)^2} {2(A-1)m}
\equiv T_{NC} - T_{NC}^{c.m.},
\end{equation}
where $T_{NC}^{c.m.}$ represents the kinetic energy due to c.m. motion of 
$(A-1)$  subsystem around the c.m. of hypernucleus. We compare the energy 
breakdown of $^4$He and $NC$ of $_\Lambda ^5$He in Table \ref{tabtwo}. A total
polarization of 3.78(4) MeV and 4.18(4) MeV is found with 
and with no $SEC$. The polarization in both cases as well as their difference
is large.  $SEC$ reduces the polarization by about 0.4 MeV.
Nemura {\it et. al.} \cite{HNemura02} also report a large polarization. 
Similar was the result for $_{\:\:\Lambda}^{17}$O \cite{UPU95}. 
We also note that the point proton 
radius in $_\Lambda^5$He with no $SEC$  is 1.66(1) fm and with $SEC$ 
is 1.62(1) fm, therefore, it is more compact with $SEC$. This is because of
the reduction in repulsive central correlation $f^c_{\Lambda N}(r)$ due to
presence of $u^x(r)$ which pushes the nucleon towards the periphery 
(cf. Fig. \ref{figtwo}).
For isolated $^4$He point proton radius is 1.46(1) fm whereas 'experimental' 
value is 1.47 fm.

The density profiles of proton ($p$) and $\Lambda$ are plotted in 
Fig. \ref{figtwo}. Like previous studies \cite{UPU95,AUsmani03}, 
most of the time $\Lambda$ is found in the interior
region. The repulsive central $f_{\Lambda N}^c$ correlation pushes the
nucleon both towards centre and at periphery. As this correlation 
gets modified
with the presence of $u^x(r)$, a  change is observed in $p$ density profile 
in $_\Lambda^5$He near the centre (upper panel) and at the peripheral 
region (lower panel). The $\Lambda$ skin is also observed in the lower panel 
where peripheral densities are plotted.

We conclude that $SEC$ is an important correlation which, being quite
significant at $r\le 2.0$ fm, modifies the $f_{\Lambda N}^c$ central 
correlation considerably at its peak and in its vicinity. 
Its effect is exhibited in energy breakdown 
of the hypernucleus, $\Lambda$-seperation energy, $NC$ ploarization 
and density profiles.  
We also note that it strongly affects the expectation value of
$V_{\Lambda NN}^{2\pi}$ and  $V^D_{\Lambda NN}$.
Therefore, it should not be ignored in any attempt to pin down the 
strenghts of $\Lambda NN$ potential. 
A detailed study to pin down these strengths and also
to resolve the $A=5$ anomaly by including
all the ground- and excited-state  s-shell single- and double-hypernuclei
is in progress.

The author acknowledges the Grant No. SP/S2/K-32/99 sanctioned to him
by the Department of Science and Technology, Government of India.   He would
like to thank S. C. Pieper for discussions that led to the conception of 
this work many years ago.

\end{document}